# Training and Operation of an Integrated Neuromorphic Network Based on Metal-Oxide Memristors

M. Prezioso[1*], F. Merrikh-Bayat[1*], B. Hoskins[1*], G. Adam[1], K. K. Likharev[2] & D. B. Strukov[1]

Despite all the progress of semiconductor integrated circuit technology, the extreme complexity of the human cerebral cortex[1], featuring in particular ~$10^{14}$ synapses, makes the hardware implementation of neuromorphic networks with a comparable number of devices exceptionally challenging. One of the most prospective candidates to provide comparable complexity, while operating much faster and with manageable power dissipation, are so-called CrossNets[2] based on hybrid CMOS/memristor circuits[3,4]. In these circuits, the usual complementary metal-oxide-semiconductor (CMOS) stack is augmented with one[3] or several[4] crossbar layers, with adjustable two-terminal resistive devices ("memristors") at each crosspoint. Recently, there was a significant progress in improvement of technology of fabrication of such memristive crossbars and their integration with CMOS circuits[5-12], including first demonstrations[5,6,12] of their vertical integration. Separately, there have been several demonstrations of discrete memristors as artificial synapses for neuromorphic networks.[13-18] Very recently such experiments were extended[19] to crossbar arrays of phase-change memristive devices. The adjustment of such devices, however, requires an additional transistor at each crosspoint, and hence the prospects of their scaling are much less impressive than those of metal-oxide memristors[20,21], whose nonlinear *I-V* curves enable transistor-free operation. Here we report the first experimental implementation of a transistor-free metal-oxide memristor crossbar with device variability lowered sufficiently to demonstrate a successful operation of a simple integrated neural network, a single layer-perceptron. The network could be taught *in situ* using a coarse-grain variety of the delta-rule algorithm[22] to perform the perfect classification of 3×3-pixel black/white images into 3 classes. We believe that this demonstration is an important step towards the implementation of much larger and more complex memristive neuromorphic networks.

---

[1]Department of Electrical and Computer Engineering, University of California at Santa Barbara, Santa Barbara, CA 93106. [2]Department of Physics and Astronomy, Stony Brook University, Stony Brook, NY 11794. Correspondence and requests for materials should be addressed to M. P. and D. B. S. (email: mprezioso@ece.ucsb.edu, strukov@ece.ucs.edu ). *These authors contributed equally to the work.

In a hybrid CMOS/memristor circuit, the CMOS subsystem contacts each wire, and hence can address each memristive device, of the add-on crossbar(s), using a specific "CMOL" area-distributed interface[3,4]. The basic idea of hybrid neuromorphic networks, CrossNets[2], is to use this opportunity for connecting CMOS-implemented hardware models of neuron bodies with the memristive crossbar(s), whose wires play the roles of axons and dendrites, with memristors mimicking biological synapses. The simple, two-terminal, transistor-free topology of metal-oxide memristive devices may enable CrossNets to achieve extremely high density – much higher than that of not only purely-CMOS neuromorphic networks (including those based on CMOS-modeled memristors[23,] and floating-gate[24] and ferroelectric[25] memory cells), but even their biological prototypes. For example, a CrossNets based on a hybrid CMOS/memristor circuit with 5 layers of 30-nm-pitch crossbars, 2 memristors per synapse, and $10^4$ synapses per neural cell would have an areal density of ~25 million cells per $cm^2$, i.e. higher than that in the human cerebral cortex, at comparable average connectivity[1]. Estimates show that at the same time, such CrossNets may provide comparable power efficiency, at a much higher operation speed – for example, an intercell signal transfer delay of ~0.02 ms (cf. ~10 ms in biology) at a readily manageable energy dissipation rate of ~1 W/$cm^2$.

However, the practical implementation of such networks is still very challenging, due to the specific physical mechanism of resistance change in most prospective, metal-oxide based memristor – reversible modulation of the concentration profile of oxygen vacancies[11,20,21]. On the positive side, the atomic scale of vacancy position modulation implies the possibility of memristor scaling down to few-nanometer dimensions, confirmed by recent experiments.[26,27] On the negative side, the scale makes the device-to-device reproducibility of device parameters, most importantly of the voltage required for memristor's electric forming[20,21], hard to achieve with currently used fabrication technologies. The device variability is the main reason why the only demonstrations of memristive neuromorphic networks we are aware of were based on disconnecting of each memristor from the crossbar for individual forming, using either a crossbar with external (off-chip) wires,[18] or an individual switch transistor at each crosspoint.[19] Both these approaches are incompatible with the goal of reaching the extremely high density of neuromorphic networks, discussed above.

The main goal of this work was the first experimental demonstration of a fully operational neural network based on a transistor-free metal-oxide memristive crossbar.

## Memristor Fabrication, Forming and Characterization

The key to the success of this work was a significant reduction of memristor variability using binary-oxide $Al_2O_3/TiO_{2-x}$ stacks (see Inset in Fig. 1b). The fabrication procedure was generally close to that described in Ref. 27, with a major difference of using low temperature (<300°C) reactive sputtering for film deposition, enabling monolithic 3D integration. The stack was optimized first of all by conducting an exhaustive experimental search over a range of titanium dioxide compositions and layer thicknesses (from 5 nm to 100 nm) to find the parameter range providing the lowest forming voltages. Within that range, the device performance, most importantly including memristor uniformity and *I-V* curve nonlinearity, was further optimized by varying the aluminum oxide thickness from 1 to 5 nm.

The main features of such optimized junctions are their low variability (Figs. S3-S4). In addition, other important characteristics of the formed 200×200 $nm^2$ devices are also very decent: the ON/OFF current ratios above 4 orders of magnitude (at ±0.1 V), a high nonlinearity of the *I-V* curves (with more than 10-fold ratio of currents at the switching voltage and a half of it), a switching endurance of at least 5000 cycles, an estimated retention of at least 10 years at room temperature, low forming (< 2V) and switching (~1.5V) voltages, and relatively low operation currents: between ~100 nA and ~100 μA – see Fig. S1.

The optimized technology was then used for the fabrication of an integrated 12×12 memristive crossbar (Fig. 1), with a few technology modifications to increase the metal electrode thickness so that the wire resistances is reduced to about 800 Ω for the top lines and 600 Ω for the bottom lines of the crossbar. The crossbars retained the excellent uniformity of virgin (pre-formed) crossbar-integrated devices (see Figs. S3-S4), allowing the individual electric forming and tuning of each memristor. The forming was performed by grounding the corresponding bottom wire and applying a current-controlled ramp-up of the top wire, while leaving all other wire potentials floating (Fig. S1a). In order to minimize current leakage during forming next devices, each formed memristor was immediately switched into its low-current (“OFF”) state. The measured individual characteristics of the formed memristors were mostly similar to those of stand-alone devices, besides a somewhat smaller (< 100) ON/OFF current ratio. This difference may be partly explained by current leakage through other crosspoints at measurements, and partly by the somewhat smaller switching voltages used for the crossbar to lower the risk of device damage. In addition, some deviations from the optimal device performance could be caused by the e-beam evaporation of thicker electrodes, which required breaking the vacuum, as opposed to fully *in-situ* sputtering of single device layers, and subsequent annealing – see Methods Summary.

## Pattern Classification Experiment

The fabricated memristive crossbar was used to implement a simple artificial neural network with the top-level (functional) scheme shown in Fig. 2a. This is a single-layer perceptron[22] with 10 inputs and 3 outputs, fully connected with 10×3 = 30 synaptic weights (Fig. 2a). As the scheme shows, perceptron's outputs $f_i$ (with $i$ = 1, 2, 3) are calculated as "activation" functions,

$$f_i = \tanh(\beta I_i), \quad (1)$$

of the vector-by-matrix product components

$$I_i = \sum_{j=0}^{9} W_{ij} V_j \ . \quad (2)$$

Here $V_j$ with $j$ = 1,…9 are the input signals, $V_0$ is a constant bias, and coefficients $W_{ij}$ are adjustable (trainable) synaptic weights. Such network is sufficient for performing, for example, a classification of 3×3-pixel black-and-white images into 3 classes, with 9 network inputs ($V_1$,…,$V_9$) corresponding to pixel values. We have tested the network on a set of $N$ = 30 patterns including 3 stylized letters ("**z**", "**v**", and "**n**") and 3 sets of 9 noisy versions of each letter, formed by flipping one of the pixels in the original image - see Fig. 2b. Because of a very limited size of the set, it was used for both training and testing.

Physically, each input signal was represented by voltage $V_j$ equal to either +0.1 V or -0.1 V corresponding, respectively, to the black or white pixel, while the bias input $V_0$ was equal to -0.1 V. Such coding makes the benchmark input set balanced, with zero average of all input signals. In order to sustain such balancing at network's output as well, each synapse was implemented with 2 memristors, so that the total number of memristors in the crossbar was 30×2

= 60. Using external electronics to enforce the virtual ground conditions on each column wire, and to subtract currents flowing in the adjacent half-columns to form a differential output signal $I_i$, we ensured that Ohm's law applied to each column of the crossbar gave the result identical to Eq. (2), with differential weights

$$W_{ij} = G_{ij}^{+} - G_{ij}^{-}, \tag{3}$$

where $G_{ij}^{\pm}$ is an effective conductance of each memristor, namely the ratio *I*/*V* at voltage 0.1 V. For our devices, these effective conductances were in the range from 10 to 100 μS, so that currents $I_i$ were of the order of a few μA. Activation functions (1) were also implemented using external electronics, with the slope $\beta = 2\times10^4$ $A^{-1}$ chosen following the recommendation in Ref. 28.

The network was trained *in situ*, i.e. without using its external computer model, by the so-called Manhattan Update rule[29], which is essentially a coarse-grain, batch-mode variation of the usual delta rule of supervised training[22]. At each iteration ("epoch") of this procedure (sketched in Fig. 2d), patterns from training set are applied, one by one, to network's input, and its outputs $f_i(n)$, where $n$ is pattern's number, are used to calculate the delta-rule weight increments

$$\Delta_{ij}(n) = e_i(n)V_j(n), \qquad \text{with } e_i(n) = \left[f_i^{(g)}(n) - f_i(n)\right] \left.\frac{df}{dI}\right|_{I=I_i(n)} \tag{4}$$

being proportional to $i$-th output error. Here $f_i^{(g)}(n)$ is the target value of $i$-th output for $n$-th input pattern. (In our system these values were accepted to equal +0.85 for the output corresponding to the correct pattern class, and -0.85 for the output corresponding to the wrong class.) Once all $N$ patterns of the training set had been applied, and all $\Delta_{ij}(n)$ calculated, the synaptic weights were modified using the following Manhattan Update rule:

$$\Delta W_{ij} = \eta \operatorname{sgn} \sum_{n=1}^{N} \Delta_{ij}(n), \tag{5}$$

where $\eta$ is a constant that scales the training rate.

Physically, in our system the weights were modified in parallel for each half-column of the crossbar (corresponding to a certain value of index $i$ in the above formulas), using two sequential voltage pulses. Namely, first a "set" pulse with amplitude $V_w^+ = 1.3$ V was applied to increase conductances of the synapses whose $\Delta G$ calculated from Eq. (5) had been positive; then a "reset" pulse $V_W^- = -1.3V$ was applied to the remaining synapses of that half-column – see Fig. 3c. This fixed-amplitude pulse procedure followed the Manhattan Update rule only approximately, because the actual training rate $\Delta G$ depends on the initial conductance $G$ of the memristor – see Fig. 1c. (For $G$ = 20 μS, $\Delta G$ was close to +60 μS for the set pulse and -5 μS for the reset pulse, while for $G$ = 65 μS, the changes were close, respectively, to +24 μS and -55 μS.) Due to this specific (though quite representative[11]) switching dynamics of our devices, the best classification performance was achieved when the memristors had been initialized somewhere in the middle of their conductance range, around 35 μS (Fig. S7b). At such initialization, the perfect classification was reached, on the average, after 15 training epochs – see Fig. 4.

In summary, we have experimentally demonstrated, for the first time, an artificial neural network using memristors integrated into a dense, transistor-free crossbar circuit. This crossbar performed, on the physical (Ohm's-law) level the analog vector-by-matrix multiplication (2)-(3), which is by far the most computationally intensive part of operation of any neuromorphic network. The other operations, described by Eqs. (1), (4), and (5), were performed by external electronics, but they are much less critical for network performance, and in future, larger CrossNets would be delegated to sparser CMOS subsystems. We believe that this demonstration is an important step toward the effective analog-hardware implementation of much more complex neuromorphic networks – first of all, multilayer-perceptron classifiers with deep learning[30], and eventually also much more elaborate CrossNet-based cognitive systems.

## Methods Summary

Crossbar lines, 200 nm wide and separated by 400 nm gaps, were formed on 4" silicon wafers covered by 200 nm of thermal $SiO_2$. After standard cleaning and rinse, fabrication started with an e-beam evaporation of Ta (5 nm)/Pt (60 nm) bilayer over a patterned photoresist (PR) to form the bottom wires. After liftoff, the wafer was descum by active oxygen dry etching at 200°C for 10 minutes. Then, a blanket film consisting of a 4-nm sputtered $Al_2O_3$ barrier and a 30-nm $TiO_2$ switching layer was deposited from a fully oxidized target and partially oxidized target, respectively. This bilayer was then removed by etching in an ICP chamber using $CHF_3$ plasma, while preserving it in the future crossbar area by pre-deposited negative photoresist.

After stripping the photoresist in the 1165 solvent for 3h at 80°C, the wafer was cleaned using a mild descum procedure performed in a RIE chamber for 15 minutes with 10 mTorr oxygen plasma at 300 W. Next, the top electrode consisting of 15 nm Ti and 60 nm Pt was deposited by e-beam evaporation; then top wires were patterned by liftoff process. Finally, the wire bonding pads were formed by e-beam deposition of Cr (10 nm) /Ni (30 nm) /Au (500nm). All lithographic steps were performed using a DUV stepper using a 248 nm laser. After fabrication and dicing, the dies were annealed in a reducing atmosphere (10% $H_2$, 90% $N_2$) for 30 minutes at 300C and wire-bonded to a DIP40 package. The final crossbar layout is shown in Figs. 1a and 2S.

All electrical characterizations were performed using the Agilent B1500A parameter analyzer. In addition, the Agilent B5250A switching matrix was employed for testing packaged crossbar circuits and carrying out the pattern classification experiment. The parameter analyzer and switching matrix were controlled by a personal computer via a GPIB interface using a custom C code. All write and read pulses were 500 μs long. For memristor adjustment we used the “$V$/2 scheme”, in which the selected row and column wires are voltage-biased at, respectively, $\pm V/2$. For device state readout, we voltage-biased the selected column wire, connected the selected row wire to a virtual ground, and physically grounded all the other lines.

## Acknowledgments

We acknowledge useful discussions with F. Alibart, I. Kataeva, W. Lu, L. Sengupta, S. Stemmer, and E. Zamanidoost. This work was supported by the Air Force Office of Scientific Research (AFOSR) under the MURI grant FA9550-12-1-0038, DARPA under Contract No. HR0011-13-C-0051UPSIDE via BAE Systems, and DENSO Corporation, Japan.

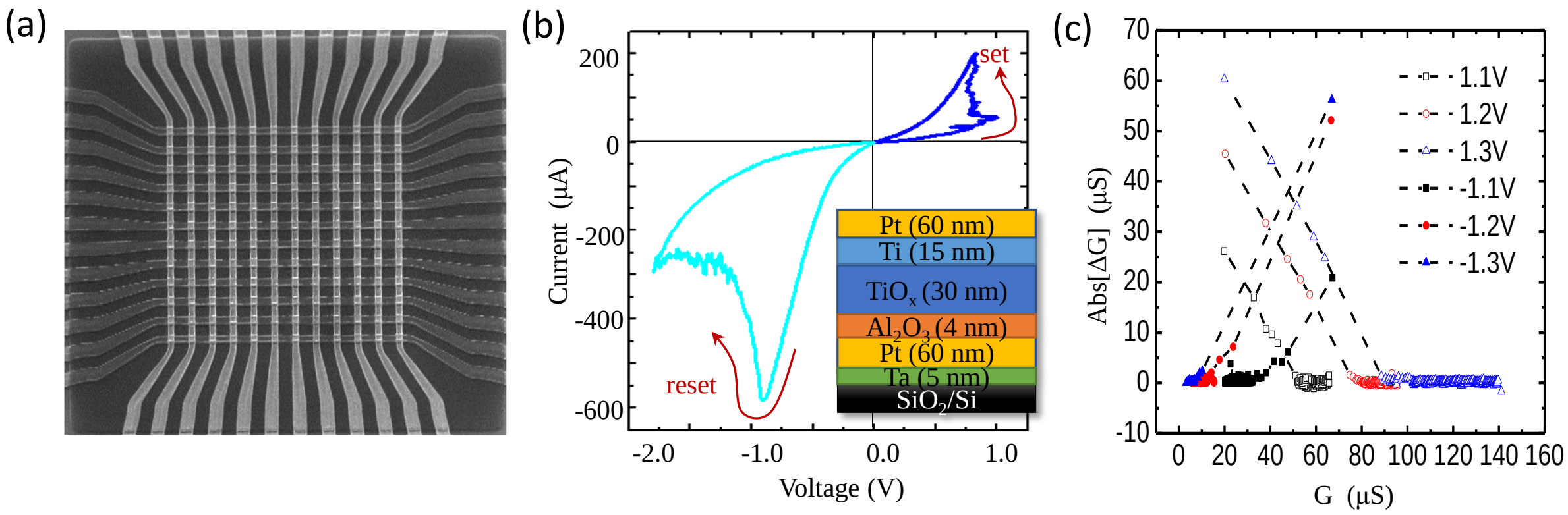

**Figure 1**. (a) Integrated 12×12 crossbar with an $Al_2O_3/TiO_{2-x}$ memristor at each crosspoint; (b) a typical *I*-*V* curve of a formed memristor, and (c) its gradual switching under the effect of 500-µs voltage pulses of two polarities, as a function of the initial conductance, for various pulse amplitudes. The inset in panel (b) shows device’s cross-section.

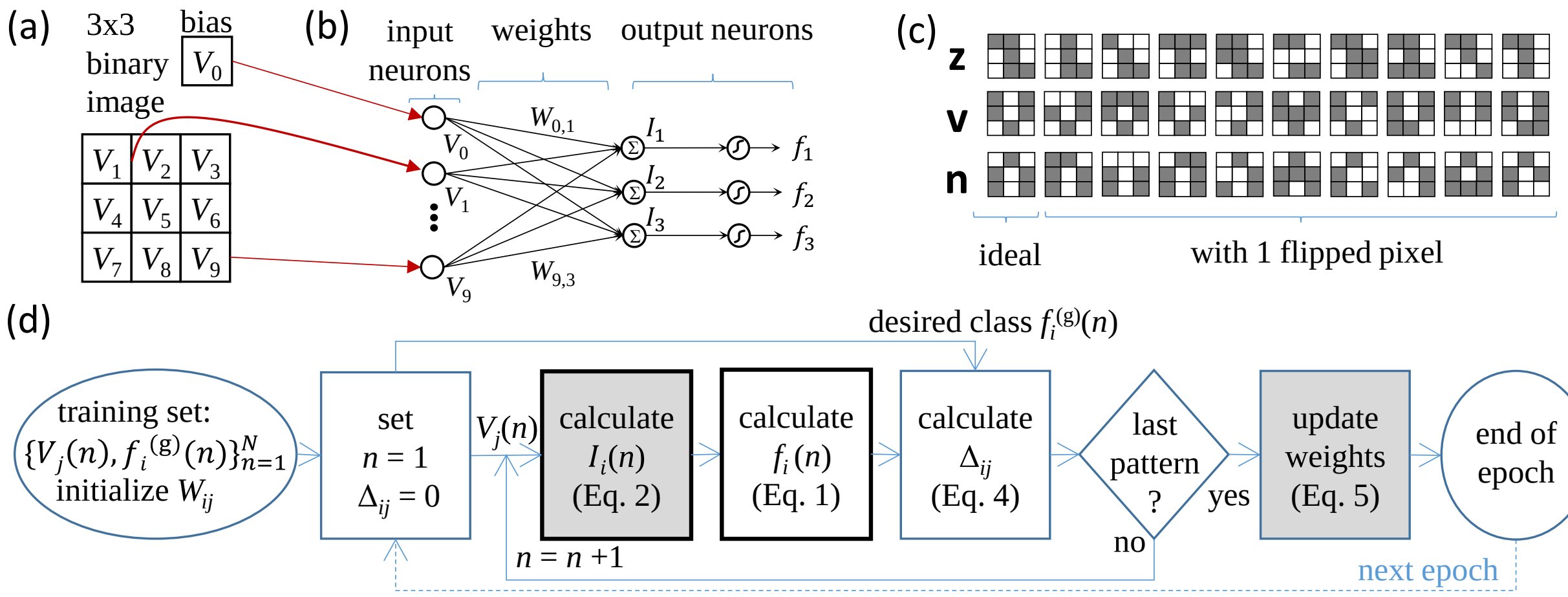

**Figure 2.** Pattern classification experiment: a top-level description. (a) Input image, (b) the single-layer perceptron for the classification of 3×3 binary images, (c) the used input pattern set, and (d) the flow chart of one epoch of our in-situ training algorithm. In panel (d), the gray-shaded boxes show the steps implemented inside the crossbar, while those with solid black borders show the only steps required for performing the classification operation.

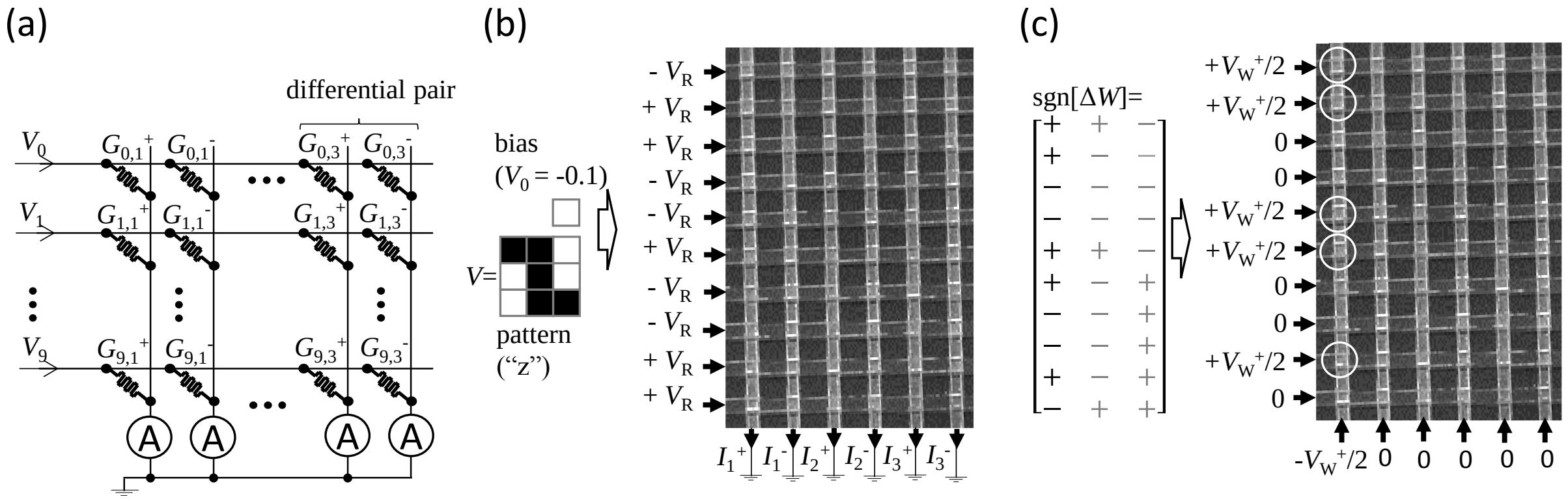

**Figure 3.** Pattern classification experiment: a physical-level description. (a) Single-layer perceptron's implementation using a 10×6 fragment of the memristive crossbar. (b) Example of the classification (i.e. signal feedforward) operation for a specific input pattern. (c) Example of the weight adjustment in one half-column (using error backpropagation) for a specific error matrix. The specific read and write biases are always $V_R$ = 0.1 V, $V_W^+$ = 1.3 V and $V_W^-$ = -1.3 V.

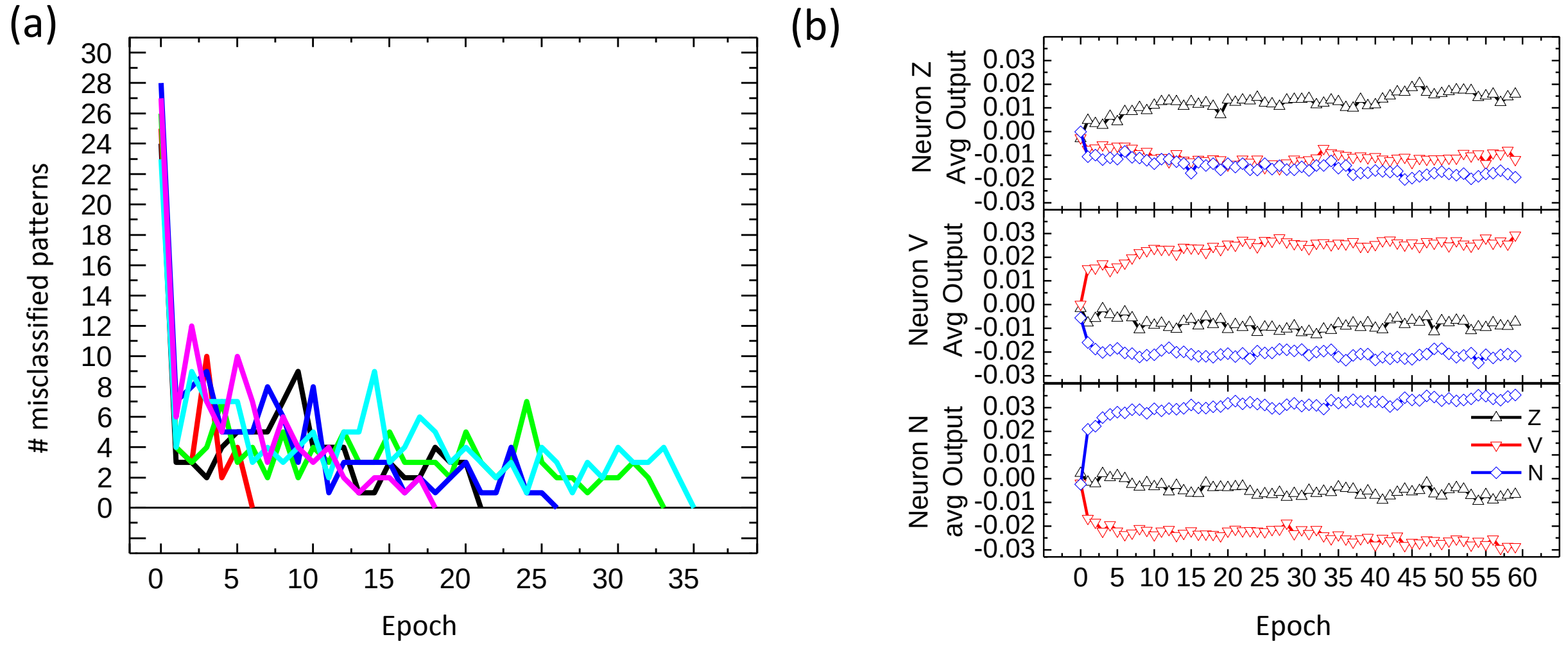

**Figure 4.** Pattern classification experiment: results. (a) Convergence of network outputs, in the process of its training, to the perfect value (zero), for 6 different initial states. (b) The evolution of output signals, averaged over all patterns of a specific class. In panel (a), the classification is considered successful when the output signal $f_i$ corresponding to the correct class of the applied pattern is larger than all other outputs.

# Supplementary Information

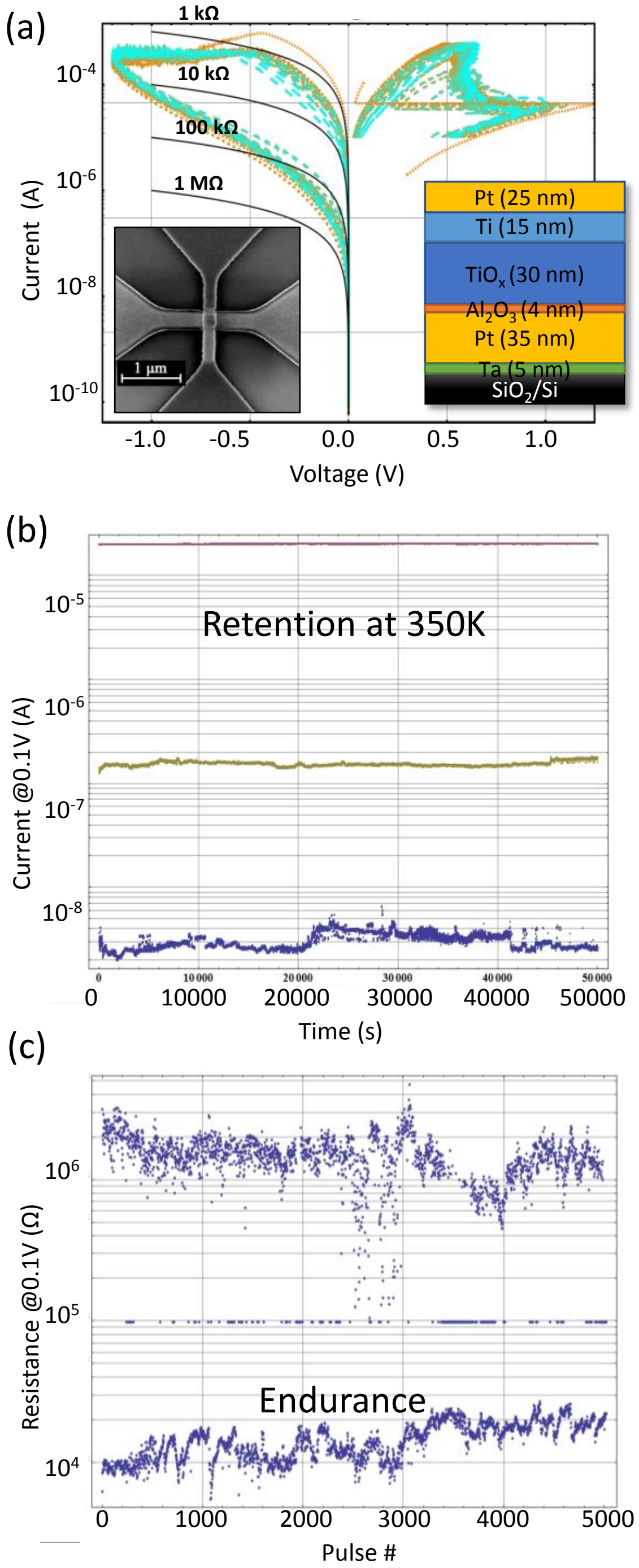

**Figure S1.** (a) A typical single Pt/$Al_2O_3$/$TiO_{2-x}$/Pt memristor and its (b) state retention and (c) switching endurance.

(a)

(b)

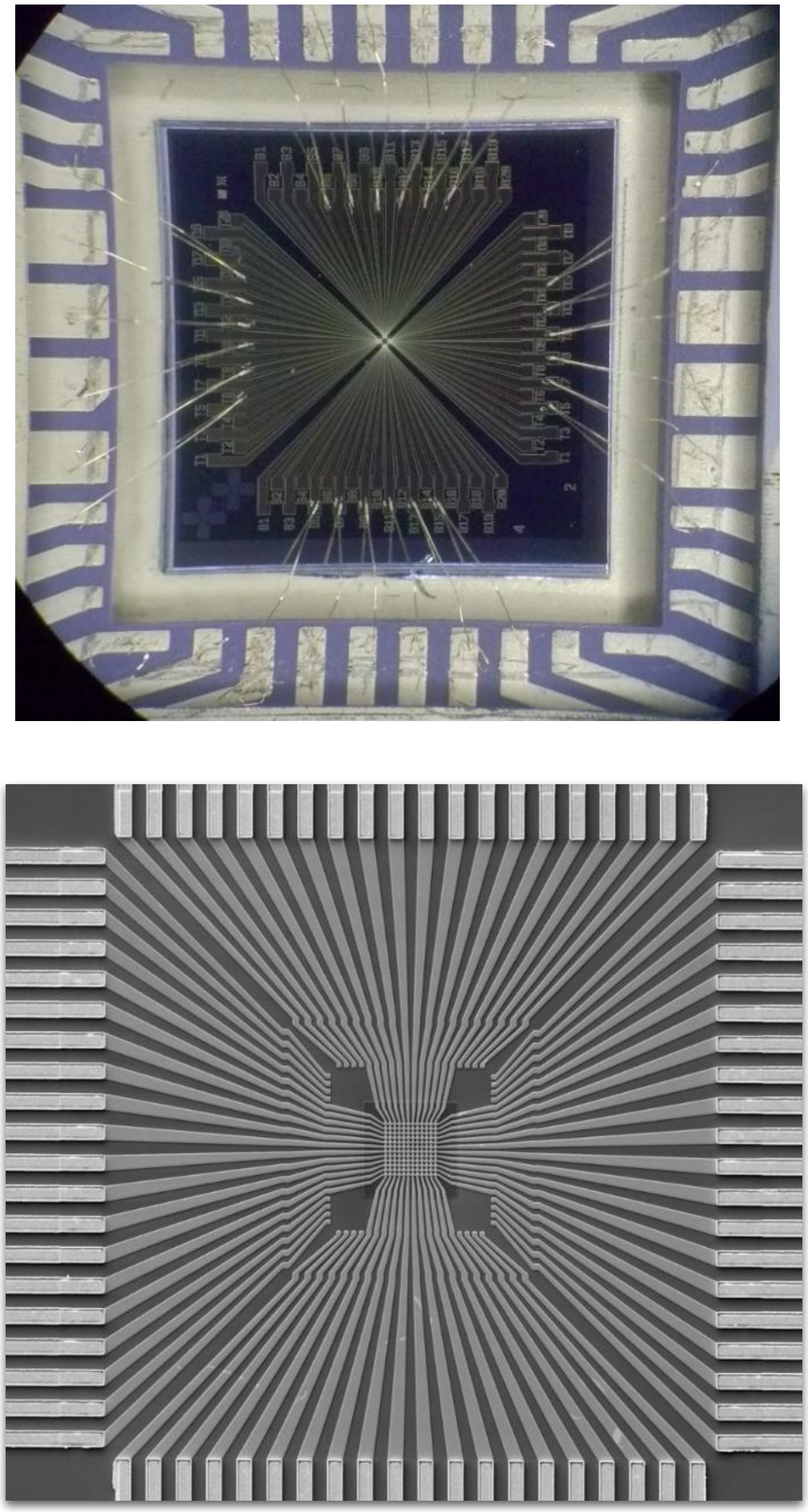

**Figure S2**. Microphotographs of the 12×12 crossbar with integrated $Pt/Al_2O_3/TiO_{2-x}/Pt$ memristive devices: (a) the bonded chip and (b) zoom-in on the crossbar area. Figure 1 in the main text shows the further zoom-in on the crossbar.

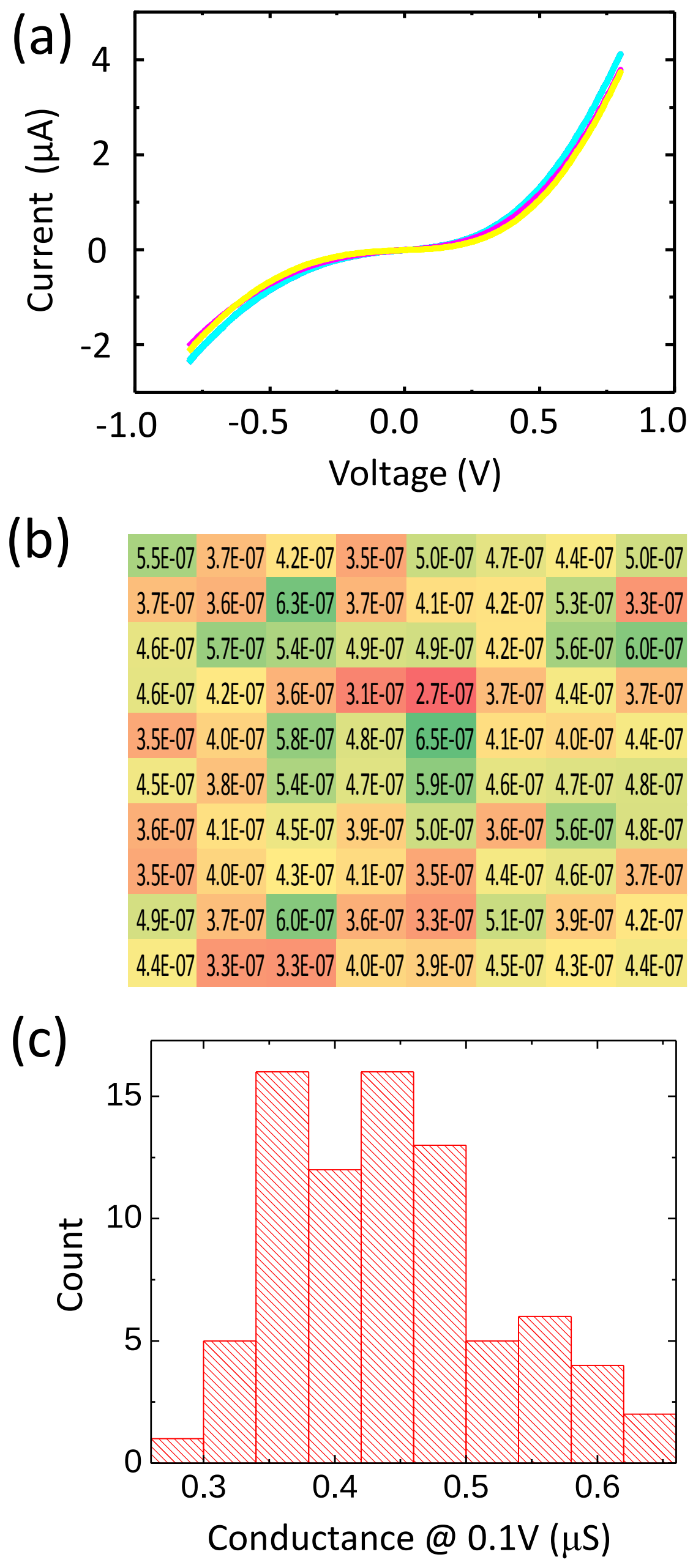

**Figure S3.** Pre-forming (virgin sample) characterization of a 10×8 portion of the crossbar: (a) Representative *I-V* curves of several devices. (b) A color-coded map and (c) histogram of device effective conductances measured at 0.1V.

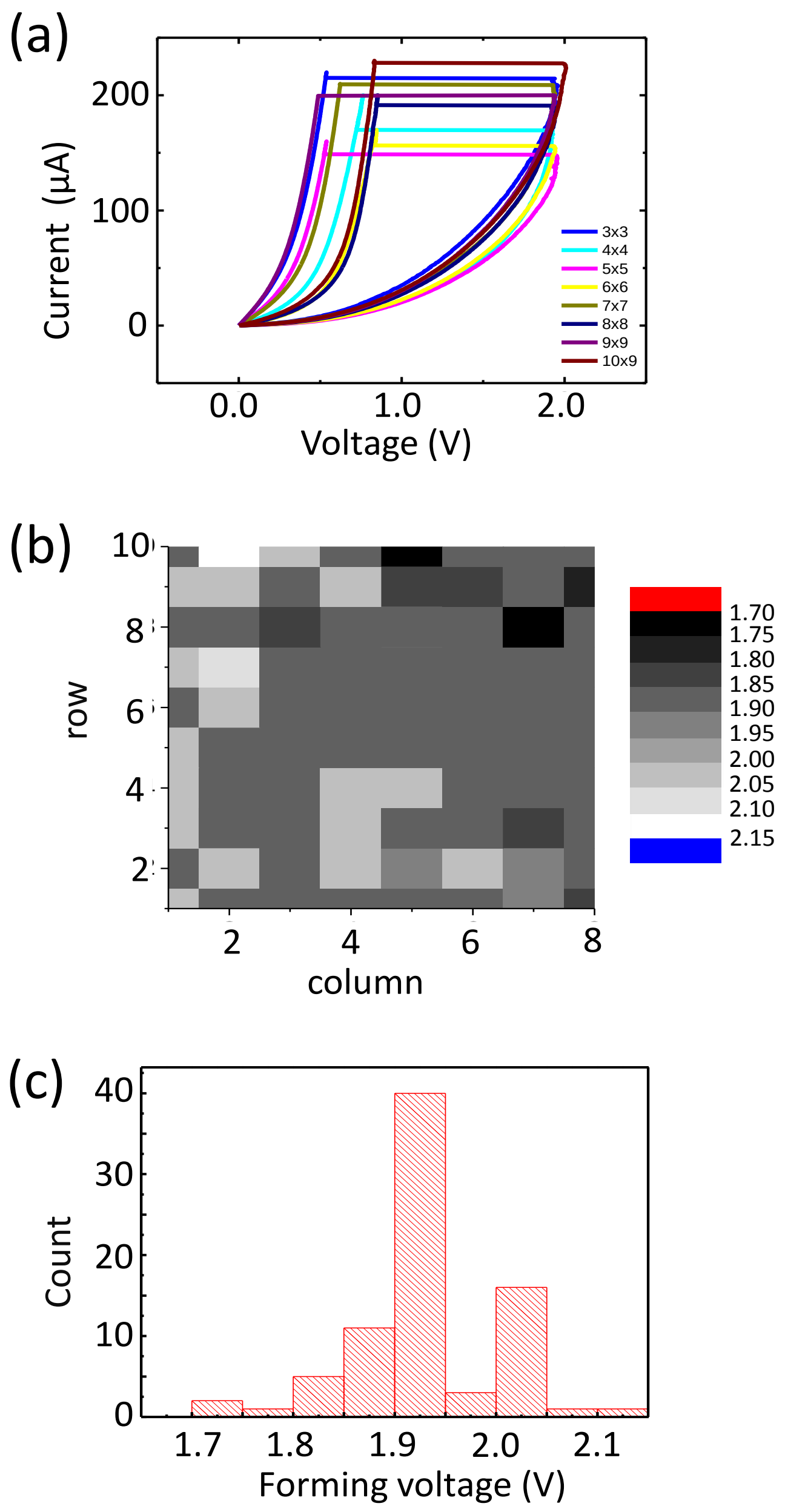

**Figure S4.** Characterization of forming in a 10×8 portion of the memristive crossbar: (a) Typical forming switching curves (b) A color-coded map and (c) histogram of forming voltages. In particular, devices were formed one-by-one first in a 2×2 array. More devices are then formed to gradually increase the size of the array. Panel a shows *I*-*V* curve of the last (typically located at the diagonal) device for a particular size of the formed array.

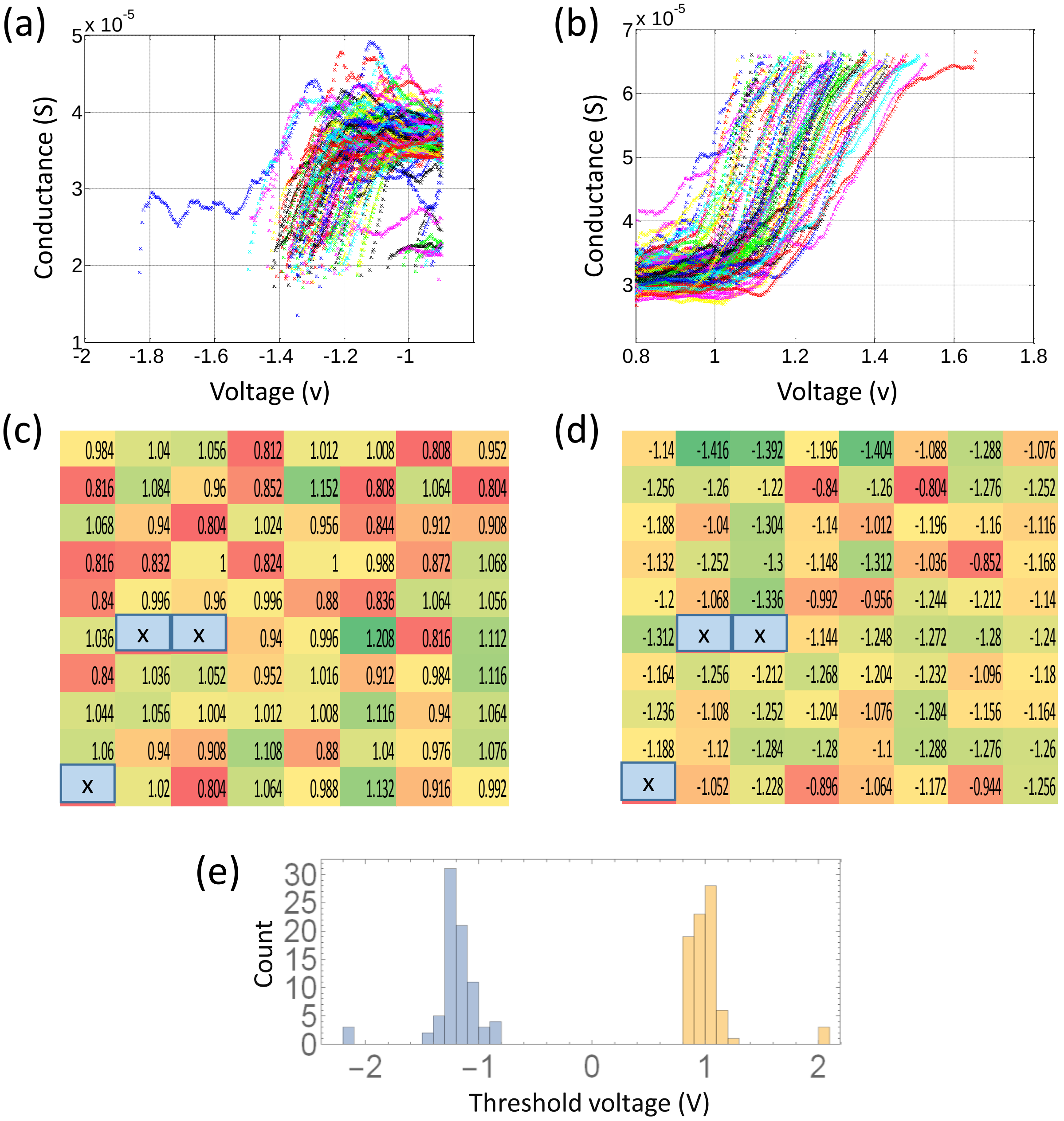

**Figure S5.** Characterization of the set and reset thresholds of formed memristors: (a, b) Conductance change as a result of an applied voltage ramps of opposite polarities and (c, d) the effective threshold voltage map for set and reset transitions, and (e) their histogram. The threshold voltages for set / reset are obtained by first programming devices to high / low resistive state and then applying a sequence of 500-μs pulses of the appropriate polarity with gradually increasing amplitude to measure the smallest voltage that causes resistance change by 2kΩ. The histogram on panel (e) does not include 3 devices which could not be switched with voltages $|V_w| \leq 1.5$V; these devices are marked with crosses on panels (c) and (d).

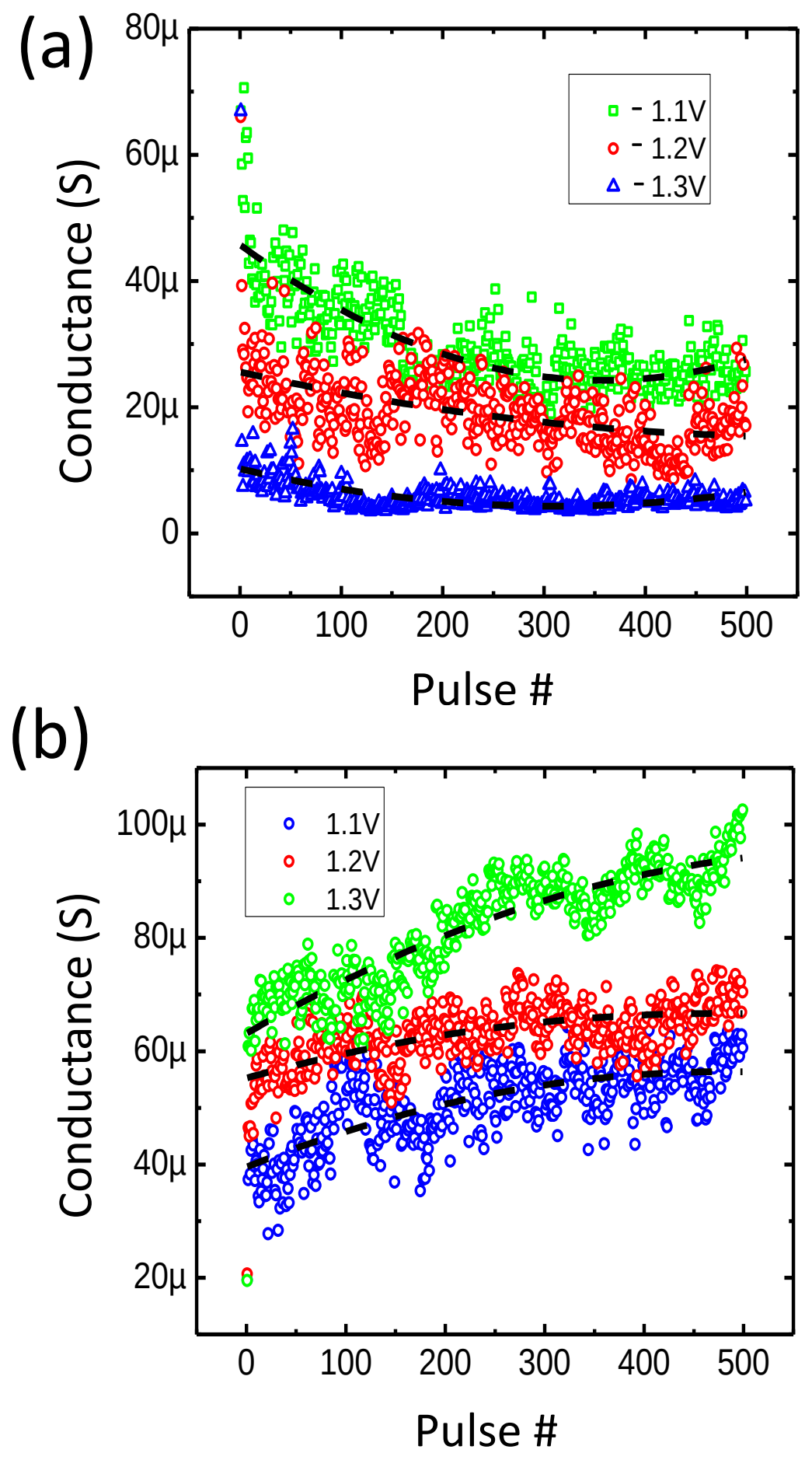

**Figure S6.** Evolution of memristor's effective conductance *I*/*V* (as measured at 0.1 V) under the effect of 500-µS pulse trains of several magnitudes, for the (a) "reset" and (b) "set" polarities. Note that unlike Figures S5a and b, this figure shows the change in conductance for a fixed amplitude pulses repeatedly applied to the same device.

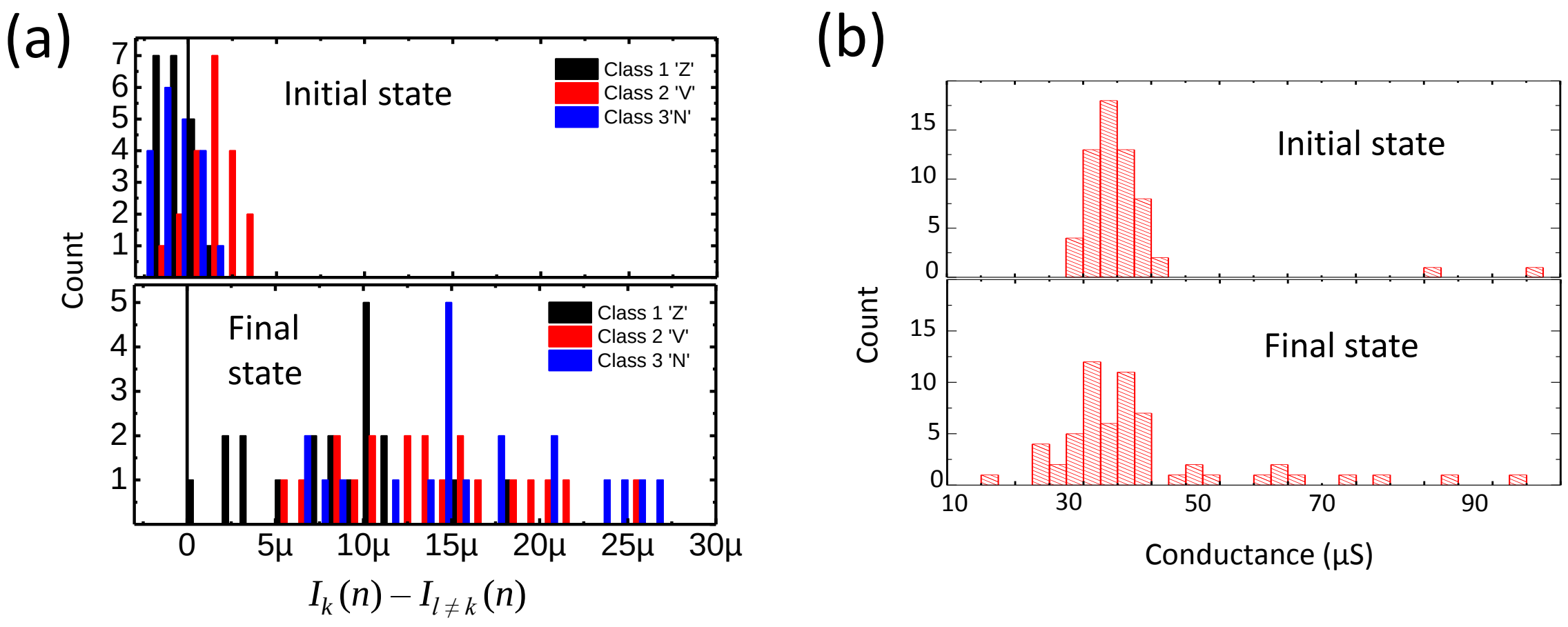

**Figure S7.** Histograms of the initial and final (a) output currents $I_i$, and (b) weights $W_{ij}$. In panel (a) index $k$ = 1,2,3 corresponds to a correct class for a given pattern $n$, so that $I_k(n)$ is a current measured at the neuron classifying pattern $n$. Alternatively, index $l$ =1, 2, 3 and currents $I_{l \neq k}(n)$ correspond to incorrect classes and currents measured at other neurons for a given class $n$, respectively. Note that the total number of data points $I_k(n) - I_{l \neq k}(n)$ for a particular class is 10×2 = 20, while it is 60 for all 3 classes.

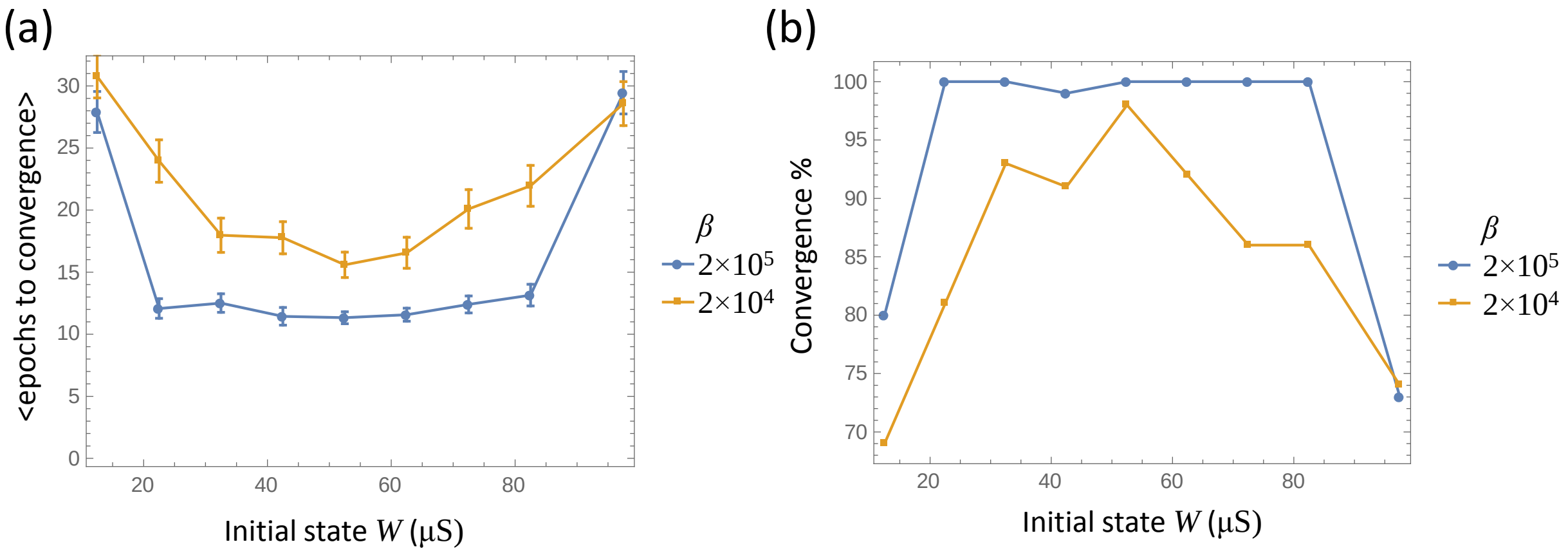

**Figure S8:** Computer simulation of the training iteration convergence as a function of weight initiation: (a) the number of epochs needed to successfully classify all patterns and (b) the percentage of experiments with perfect convergence. Each point is an average over 100 runs, each with the memristors randomly initialized within a 5-μS conductance window around the value on the horizontal axis. The maximum number of epochs is set to 50 so that if an experiment takes more than 50 epochs to convergence it is considered as a failure.